\documentclass[%
reprint,
 amsmath,amssymb,
 aps,
]{revtex4-1}

\usepackage{graphicx}
\usepackage{dcolumn}
\usepackage{bm}
\usepackage{amsmath}

\begin{document}
\title{A numerical investigation on the convergence issues for ghost imaging}%

\author{Minghui Zhang}
\email{zmh@ahu.edu.cn}
\affiliation{School of Physics and Material Science, Anhui University, No.111
Jiulong Road, Hefei, 230601, China}

\date{\today}

\begin{abstract}
The long time consumption is a bottleneck for the applicability of the ghost imaging (GI). By introducing a criterion for the convergence of GI, we investigate a factor that impacts on the convergence speed of it. Based on computer experiments, we demonstrate that the object's feature size relative to the spatial coherent length of the illuminating light impacts on the necessary number of uncorrelated data being acquired for the correlation computation. It may motivate people to seek ways towards real-time practical applications of GI and its analogues. In addition, the method to simulate the uncorrelated sequence of a complex ensemble for thermal light is valuable in applications where actively controlling the light fields is needed.
\end{abstract}


\maketitle

\section{\label{sec:1}Introduction}
In 1994, A. V. Bilinski\v{\i} and D.N. Klyshko found that momentum entangled two-photon pairs generated by the parametric downcoversion should experience a distinctive phenomenon \cite{1}. In their prediction, an object's diffraction pattern was obtained by counting the coincidence rate. Because it was a function of spatial variable in the optical path that actually never passed the object, the obtained patterns were called ghost imaging or ghost interference (GI). From then on, the GI has been undergoing excessive investigations. Although debate issues on its quantum and classical essence exist \cite{2,3}, it has been accepted that ghost imaging can be achieved by both quantum entangled two-photon pairs and classical thermal light \cite{4,5,6,7,8,9,10}. Nowadays, Applications for GI span over a wide area in experimental optics with examples on quantum lithography \cite{11}, supper resolution imaging \cite{12}, coherent x-ray diffraction imaging \cite{8}, phase object determination \cite{13,13,15,16}, holography \cite{17}, and imaging through atmospheric turbulence \cite{18,19}, to name a few. As a GI's analogue approach \cite{20}, the computational imaging with single-pixel has also been reported in resent years \cite{21,22,23}.

To achieve GI (and its analogues), the iterative algorithm based correlation computation is an indispensable work. It requires a large number $(N)$ of uncorrelated offline data. In practice, the more precisely the ghost imaging is to convergent, the larger sample number of $N$ is needed for the iteration. One may get a glimpse of the convergent process of this kind from Fig.$2$ of Ref.\cite{23}. Since the value of $N$ dominates the time consumption for the data acquisition, obviously, one must consider the time consumption issues \cite{Wu}, especially on the occasion of imaging those dynamical or ephemeral objects. Otherwise the GI's application value will be largely discounted. However, the topic  for finding ways to yield a satisfied precisely convergent GI with minimal $N$ has not been reported quantitatively yet.

In order to valuate the convergent quality, we introduce a criterion based on the concept known as ``root-mean standard error",
\begin{align}
{\varepsilon _{RM}} \equiv \sqrt {\frac{{\sum\limits_{i = m}^n {{{\left( {{{\hat y}_i} - {y_i}} \right)}^2}} }}{{n - m + 1}}},
\label{eq:1}
\end{align}
between the image (``experimental data" hereinafter) and the imaging target (``theoretical prediction" hereinafter). In Eq. \ref{eq:1}, the positive integers $m < n$ ; the $\hat{y}_{i}$  and the  $y_{i}$  are, respectively, the GI's experimental data and the theoretical prediction of the $i^{th}$ pixel ($m \leqslant i \leqslant n$), both of them had been normalized within an interval of $[0, 1]$ to stand for the gray scale. We have already introduced that  $\hat{y}_{i}$ converges to $y_{i}$ as $N$ increases. Clearly once an appreciable value of the $\varepsilon_{RM}$ , say $0.07$, reaches, one may regard that the $N$ is large enough to accomplish a sound ghost imaging. Of cause the somewhat arbitrary choice of $\varepsilon_{RM}  = 0.07 $ may be replaced by other values, depending on the purpose. In this letter, by computer experiments, we applied this criterion to investigate a factor that will impact on the relation between $\varepsilon_{RM}$ and $N$.

We give a necessary review on GI in Sec. \ref{sec:2}. After that, the preparations for  computer experiments are described in Sec. \ref{sec:3}. Then, in Sec. \ref{sec:4}, four computer experiments with their results are presented. Following that, we make conclusions in Sec. \ref{sec:5}.

\section{\label{sec:2}A framework of theoretical bases and experimental scheme}

\begin{figure}[htb]\includegraphics
[bb=11 13 234 159,scale=1.1]
{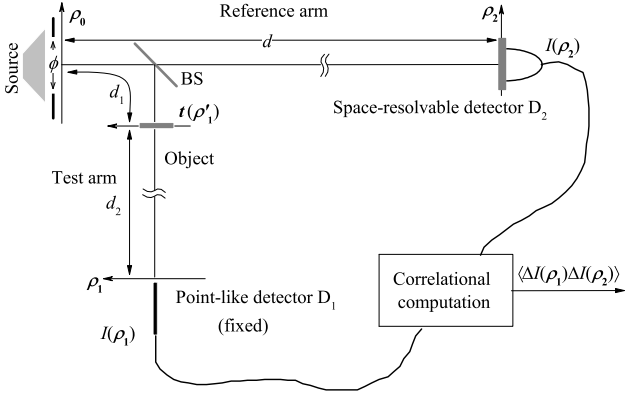}\caption{\label{fig:1} (Color online)  The set up scheme for ghost interference.}
\end{figure}

We only review the scheme for ghost interference as Fig. \ref{fig:1} shows, because its analogues to ghost geometry imaging and single pixel imaging are straightforward. In the setup we suppose that the thermal source is limited by a disk like aperture with a diameter of $\phi$. The vectors $\boldsymbol{\rho_{0}}$, $\boldsymbol{\rho_{1}}$, $\boldsymbol{\rho_{2}}$, and $\boldsymbol{\rho'_{1}}$ in Fig. \ref{fig:1} denote position variables, respectively, for the thermal source fields, detectors $D_{1}$, $D_{2}$, and the object .The optical field $\boldsymbol{E}(\boldsymbol{\rho_{0}})$ from the source was split into a pair of twin copies by a $50/50$ beam splitter  to form two optical paths. One of them being called the test arm contains an object with the amplitude transmittance of $\boldsymbol{t}(\boldsymbol{\rho'_{1}})$ at a
distance of $d_{1}$ from the source. After the object with a distance of $d_{2}$, a time varying intensities $I(\boldsymbol{\rho_{1}})$ of a fluctuating light field $\boldsymbol{E}(\boldsymbol{\rho_{1}})$ was collected by a fixed point-like detector $D_{1}$, which is spatially none resolvable. The other optical path is being called reference arm. It contains no object. In the reference arm, after a free space with a distance of $d$ from the source, a time varying intensities $I(\boldsymbol{\rho_{2}})$ of a fluctuating light field $\boldsymbol{E}(\boldsymbol{\rho_{2}})$ was collected by a detector $D_{2}$. Unlike $D_{1}$, the detector $D_{2}$ is a spatially resolvable one. It had been theoretically and experimentally approved \cite{7,8}, when
\begin{align}
d = {d_1} + {d_2},
\label{eq:2}
\end{align}
the correlation function of fluctuating parts of $I(\boldsymbol{\rho_{1}})$ and $I(\boldsymbol{\rho_{2}})$ recovers the object $\boldsymbol{t}(\boldsymbol{\rho'_{1}})$ in Fourier space in a way of
\begin{align}
G\left( {{{\boldsymbol{\rho }}_{\boldsymbol{2}}},{{\boldsymbol{\rho }}_{\boldsymbol{1}}}} \right) \equiv \left\langle {\Delta I\left( {{{\boldsymbol{\rho }}_{\boldsymbol{2}}}} \right)\Delta I\left( {{{\boldsymbol{\rho }}_{\boldsymbol{1}}}} \right)} \right\rangle  = {\left| {{\boldsymbol{T}}\left( {\frac{{{{\boldsymbol{\rho }}_{\boldsymbol{2}}} - {{\boldsymbol{\rho }}_{\boldsymbol{1}}}}}{{\lambda {d_2}}}} \right)} \right|^2};
\label{eq:3}
\end{align}
where angle brackets denotes ensemble average; in the equation, $\Delta{I}(\boldsymbol{\rho_{k}}) \equiv I(\boldsymbol{\rho_{k}})- <I(\boldsymbol{\rho_{k}}) >$, $(k = 1,2.)$, is the fluctuating part of $I(\boldsymbol{\rho_{k}})$; $\lambda$ is the wavelength of the monochromatic thermal source, and $\boldsymbol{T}(\boldsymbol{\xi})$ is the Fourier transform of the object $\boldsymbol{t}(\boldsymbol{\rho'_{1}})$ , defined by
\begin{align}
{\boldsymbol{T}}\left( {\boldsymbol{\xi }} \right) \equiv \int_\infty  {{\boldsymbol{t}}\left( {\boldsymbol{\rho'_1}} \right){e^{ - {\boldsymbol{j}}2\pi \boldsymbol{\rho'_1}{\boldsymbol{\xi }}}}d\boldsymbol{\rho'_1}} .
\label{eq:4}
\end{align}
Because the point-like detector $D_{1}$ senses the optical fields only at a fixed position of $\boldsymbol{\rho_{1}} = 0$, the correlation function of Eq. \ref{eq:3} related only to the variable of $\boldsymbol{\rho_{2}}$, which was in the optical path that actually never pass the object. For this unique feature, the obtained pattern with a function of $\boldsymbol{\rho_{2}}$,
\begin{align}
G\left( {{{\boldsymbol{\rho }}_{\boldsymbol{1}}} = 0,{{\boldsymbol{\rho }}_{\boldsymbol{2}}}} \right) = {\left| {{\boldsymbol{T}}\left( {\frac{{{{\boldsymbol{\rho }}_{\boldsymbol{2}}}}}{{\lambda {d_2}}}} \right)} \right|^2},
\label{eq:5}
\end{align}
is being named as the ghost diffraction pattern, or the ghost imaging in Fourier space \cite{24}. In practice, the correlation function of $G$ was realized by offline computation with an iterative algorithm of
\begin{align}
\begin{array}{l}
G\left( {{{\boldsymbol{\rho }}_{\boldsymbol{1}}} = 0,{{\boldsymbol{\rho }}_{\boldsymbol{2}}}} \right) = \frac{1}{N}\sum\limits_{n = 1}^N {{I^{\left( n \right)}}\left( {{{\boldsymbol{\rho }}_{\boldsymbol{1}}} = 0} \right){I^{\left( n \right)}}\left( {{{\boldsymbol{\rho }}_{\boldsymbol{2}}}} \right)} \\
\begin{array}{*{20}{c}}
{}&{}&{}&{}&{}&{}{}&{}&{}&{}&{}&{}
\end{array}   - \frac{1}{{{N^2}}}\sum\limits_{n = 1}^N {{I^{\left( n \right)}}\left( {{{\boldsymbol{\rho }}_{\boldsymbol{1}}} = 0} \right)} \sum\limits_n^N {{I^{\left( n \right)}}\left( {{{\boldsymbol{\rho }}_{\boldsymbol{2}}}} \right),} \\
\begin{array}{*{20}{c}}
{}&{}&{}&{}&{}&{}{}&{}&{}&{}&{}&{}&{}{}&{}&{}&{}&{}&{}
\end{array}\left( {n = 1,2,...,N} \right).
\end{array}
\label{eq:6}
\end{align}
In Eq. \ref{eq:6}, $I^{(n)}(\boldsymbol{\rho_{k}})$, $(k = 1, 2; n = 1, 2,\ldots,N.)$, are the $n^{th}$ offline stored data being acquired by detectors $D_{1}$ and $D_{2}$ respectively.

By the way, it is worth to briefly introduce the mechanism of the single-pixel imaging. It is an analogue to GI \cite{20}. The main changes are to actively control the values of the light source $\boldsymbol{E}(\boldsymbol{\rho_{0}})$ , [e.g. by a spatial light modulator (LSM)], to replace $I_{2}(\boldsymbol{\rho_{2}})$ by those pre-stored ones predicted from $\boldsymbol{E}(\boldsymbol{\rho_{0}})$, and, hence to remain only a single pixel to acquire data $I^{n}(\boldsymbol{\rho_{1}} = 0)$.

\section{\label{sec:3}Preparations for computer experiments}
\subsection{\label{sec:leve3.1}Theoretical bases for the algorithm}
We mainly describe the computer experiment for ghost interference based on the setup as shown by Fig. \ref{fig:1}. Its kernel is to simulate an uncorrelated sequence $\boldsymbol{E}^{(n)}(\boldsymbol{\rho_{0}})$, $(n = 1, 2, \ldots)$, of the realizations of an complex ensemble $\{\boldsymbol{E}(\boldsymbol{\rho_{0}}) = A(\boldsymbol{\rho_{0}})e^{\boldsymbol{j}\varphi(\boldsymbol{\rho_{0}})}\}$, which describes the fluctuated optical fields of the thermal source. Where, the amplitude $A$ and the phase $\varphi$ obeyed the continuous probability distributions for positive-valued random variables. It is well known that the most widely existing thermal light in nature can be modeled by a complex circular Gaussian random process with zero mean \cite{25}. This feature equals to command $A$ to obey Rayleigh distribution \cite{25},
\begin{align}
p\left( A \right) = \left\{ {\begin{array}{*{20}{c}}
{\frac{A}{{{\sigma ^2}}}{e^{ - \frac{{{A^2}}}{{2{\sigma ^2}}}}},}&{A > 0,}\\
{0,}&\operatorname{elsewhere,}
\end{array}} \right.
\label{eq:7}
\end{align}
the parameter $\sigma^{2}$ being the variance, and to command $\varphi$ to obey uniform distribution \cite{25}:
\begin{align}
p\left( \varphi  \right) = \left\{ {\begin{array}{*{20}{c}}
{\frac{1}{{2\pi }},}&{0 < \varphi  \le 2\pi ,}\\
{0,}&\operatorname{elsewhere,}
\end{array}} \right.
\label{eq:8}
\end{align}
respectively. Besides, the mutual independence of the two random variables requires
\begin{align}
p\left( {A,\varphi } \right) = p\left( A \right)p\left( \varphi  \right).
\label{eq:9}
\end{align}
In Eqs. \ref{eq:7} - \ref{eq:9}, $p$'s are probability densities.

Once $\boldsymbol{E}^{(n)}(\boldsymbol{\rho_{0}})$ was modeled, the sequences of the $n^{th}$ realization of $\boldsymbol{E}^{(n)}(\boldsymbol{\rho'_{1}})$ of ensembles $\{\boldsymbol{E}(\boldsymbol{\rho'_{1}})\}$, was determined by Fresnel integral
\begin{align}
{{\boldsymbol{E}}^{\left( n \right)}}\left( {{{\boldsymbol{\rho'_1}}}} \right) = \int {{{\boldsymbol{E}}^{\left( n \right)}}\left( {{{\boldsymbol{\rho }}_{\boldsymbol{0}}}} \right){\boldsymbol{h}}\left( {{{\boldsymbol{\rho }}_{\boldsymbol{0}}},{{\boldsymbol{\rho'_1}}}} \right){\boldsymbol{d}}{{\boldsymbol{\rho }}_{\boldsymbol{0}}}} ,\begin{array}{*{20}{c}}
{}
\end{array}\left( {n = 1,2,...} \right),
\label{eq:10}
\end{align}
in which, the propagator is
\begin{align}
{\boldsymbol{h}}\left( {{{\boldsymbol{\rho }}_{\boldsymbol{0}}},\boldsymbol{\rho'_1}} \right) = \frac{{{e^{{\boldsymbol{j}}k{d_1}}}}}{{{\boldsymbol{j}}\lambda {d_1}}}{e^{{\boldsymbol{j}}\frac{k}{{2{d_1}}}{{\left( {\boldsymbol{\rho'_1} - {{\boldsymbol{\rho }}_{\boldsymbol{0}}}} \right)}^2}}},
\label{eq:11}
\end{align}
under the paraxial approximation. In the same way, the sequences of the $n^{th}$ realization $\boldsymbol{E}^{(n)}(\boldsymbol{\rho_{1}})$  of ensembles $\{\boldsymbol{E}(\boldsymbol{\rho_{1}})\}$, was determined by
\begin{align}
{{\boldsymbol{E}}^{\left( n \right)}}\left( {{{\boldsymbol{\rho }}_{\boldsymbol{1}}}} \right) = \int {{{\boldsymbol{{ E}}}^{\left( n \right)}}\left( {\boldsymbol{\rho'_1}} \right){\boldsymbol{h}}\left( {\boldsymbol{\rho'_1},{{\boldsymbol{\rho }}_{\boldsymbol{1}}}} \right){\boldsymbol{d}}\boldsymbol{\rho'_1}} ,\begin{array}{*{20}{c}}
{}
\end{array}\left( {n = 1,2,...} \right),
\label{eq:12}
\end{align}
where, the propagator is
\begin{align}
{\boldsymbol{h}}\left( {\boldsymbol{\rho'_1},{{\boldsymbol{\rho }}_{\boldsymbol{1}}}} \right) = \frac{{{e^{{\boldsymbol{j}}k{d_2}}}}}{{{\boldsymbol{j}}\lambda {d_2}}}{e^{{\boldsymbol{j}}\frac{k}{{2{d_1}}}{{\left( {{{\boldsymbol{\rho }}_{\boldsymbol{1}}} - \boldsymbol{\rho'_1}} \right)}^2}}};
\label{eq:13}
\end{align}
and the sequences of the $n^{th}$ realization $\boldsymbol{E}^{(n)}(\boldsymbol{\rho_{2}})$ of ensembles $\{\boldsymbol{E}(\boldsymbol{\rho_{2}})\}$, was determined by
\begin{align}
{{\boldsymbol{E}}^{\left( {n} \right)}}\left( {{{\boldsymbol{\rho }}_{\boldsymbol{2}}}} \right){\boldsymbol{ = }}\int {{{\boldsymbol{E}}^{\left( {n} \right)}}\left( {{{\boldsymbol{\rho }}_{\boldsymbol{0}}}} \right){\boldsymbol{h}}\left( {{{\boldsymbol{\rho }}_{\boldsymbol{0}}}{\boldsymbol{,}}{{\boldsymbol{\rho }}_{\boldsymbol{2}}}} \right){\boldsymbol{d}}{{\boldsymbol{\rho }}_{\boldsymbol{0}}}} ,\begin{array}{*{20}{c}}
  {}
\end{array}\left( {n = 1,2,...} \right),
\label{eq:14}
\end{align}
the propagators is
\begin{align}
{\boldsymbol{h}}\left( {{{\boldsymbol{\rho }}_{\boldsymbol{0}}},{{\boldsymbol{\rho }}_{\boldsymbol{2}}}} \right) = \frac{{{e^{{\boldsymbol{j}}kd}}}}{{{\boldsymbol{j}}\lambda d}}{e^{j\frac{k}{{2d}}{{\left( {{{\boldsymbol{\rho }}_{\boldsymbol{2}}} - {{\boldsymbol{\rho }}_{\boldsymbol{0}}}} \right)}^2}}}.
\label{eq:15}
\end{align}
Before performing the experiment, one needs to change the illuminating light's spatial coherent length $l_{c}$ of the position $\boldsymbol{\rho'_1}$ of the object. This can be achieved by modulate the aperture's diameter $\phi$ of the source because $l_{c}$ is often estimated by Van Cittert-Zernike (VCZ) theorem \cite{25} in a way of:
\begin{align}
{l_c} = \lambda \frac{{{d_2}}}{\phi }.
\label{eq:16}
\end{align}
After selecting the object, arranging the setup parameters $d$, $d_{1}$, and $d_{2}$, choosing the $\sigma^{2}$, and modulating $\phi$, one can program the computer to simulate an independent sequence of  $\boldsymbol{E}^{(n)}(\boldsymbol{\rho_{0}})$, $(n = 1, 2,\ldots, N)$, according to Eqs.\ref{eq:7} - \ref{eq:9}, and thus to produce $I^{(n)} (\boldsymbol{\rho_{k}}) = |\boldsymbol{E}^{(n)}(\boldsymbol{\rho_{k}})|^{2}$, $(k = 1,2.)$, according to Eqs.\ref{eq:10} - \ref{eq:15}. Finally, the experimental results are calculated with the algorithm of Eq. \ref{eq:6} as they were done in real experiments.

\subsection{\label{sec:leve3.2}Samples related to instantaneous intensities of thermal optical fields simulated by computer}

\begin{figure*}\includegraphics
[bb=35 60 461 282,scale=1.0]
{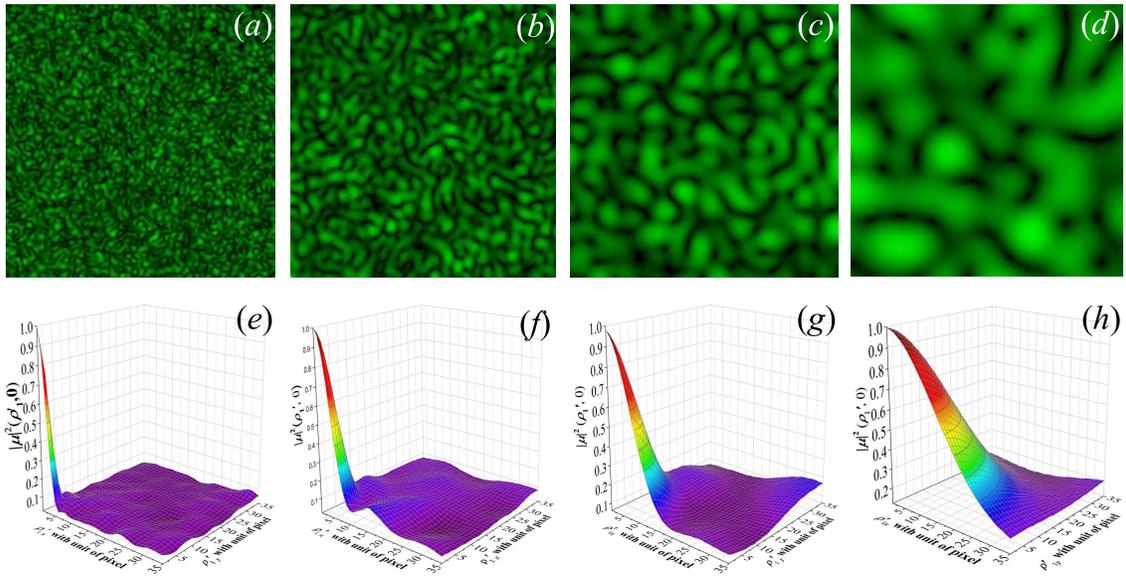}\caption{\label{fig:2} (Color online) (\emph{a})-(\emph{d}), shoots for four realizations of the instantaneous intensity ensembles of $\{I(\boldsymbol{\rho'_1})\}$ when $\phi_{k} = 2.5\times(\frac{1}{2})^{k}$ mm,($k = 0, 1, 2, 3$);they are simulated by the computer. (\emph{e})-(\emph{h}), the squired-modules of the coherence factor $|\mu(\boldsymbol{\rho'_1},0)|^{2}$, corresponding to the thermal light fields $E(\boldsymbol{\rho'_1})$ whose instantaneous intensities  have described by (\emph{a})-(\emph{d}).}
\end{figure*}

To demonstrate the validity of the computer simulated thermal optical fields, in this subsection we investigate four typical realizations of the instantaneous intensity sequencies belonging to the ensembles of $\{I(\boldsymbol{\rho'_{1}})\}$, as examples. With the statistical properties given by Eqs.\ref{eq:7} - \ref{eq:9}, we programmed the computer to simulate sequences of $\boldsymbol{E}^{(n)}(\boldsymbol{\rho_{0}})$. They are the realizations of ensemble $\{\boldsymbol{E}(\boldsymbol{\rho_{0}})\}$. In the program, the parameter of $\sigma^{2}$ is set to be unit. After choosing $d_{2} = 60$ mm and setting aperture's diameters to $\phi_{k} = 2.5\times(1/2)^{k}$ mm, $(k = 0, 1, 2, 3)$, respectively, the sequences of intensity realizations  $I^{(n)}(\boldsymbol{\rho'_{1}})$, $(n=1,2,\ldots)$, of ensemble $\{I(\boldsymbol{\rho'_{1}})= |\boldsymbol{E}(\boldsymbol{\rho'_{1}})|^{2}\} $ were produced according to Eqs.\ref{eq:10} and \ref{eq:11}. Their typical shoots for those four realizations are shown by Fig. \ref{fig:2} (\emph{a}) - (\emph{d}) respectively. We can see each one of them features like a laser speckle pattern \cite{26}, and the size of those speckles increases while $\phi_{k}$ decreases.They are consistent with the physical picture of the instantaneous intensities for the monochromatic thermal light \cite{add_before_27}.

For those four sample realizations, we calculated the squired-modules of their coherence factor $|\mu(\boldsymbol{\rho'_{1}},0)|^{2}$ \cite{27} in a way of
\begin{widetext}
\begin{align}
{\left| {\mu \left( {\boldsymbol{\rho'_1},0} \right)} \right|^2} = \frac{{\frac{1}{N}\sum\limits_{n = 1}^N {{I^{\left( n \right)}}\left( {\boldsymbol{\rho'_1}} \right){I^{\left( n \right)}}\left( {\boldsymbol{\rho'_1} = 0} \right)}  - \frac{1}{N^2}\sum\limits_{n = 1}^N {{I^{\left( n \right)}}\left( {\boldsymbol{\rho'_1}} \right)\sum\limits_{n = 1}^N {{I^{\left( n \right)}}\left( {\boldsymbol{\rho'_1} = 0} \right)} } }}{{\frac{1}{{{N^2}}}\sum\limits_{n = 1}^N {{I^{\left( n \right)}}\left( {\boldsymbol{\rho'_1}} \right)\sum\limits_{n = 1}^N {{I^{\left( n \right)}}\left( {\boldsymbol{\rho'_1} = 0} \right)} } }},\begin{array}{*{20}{c}}
  {}
\end{array}\left( {n = 1,2,...,N} \right).
\label{eq:17}
\end{align}
\end{widetext}
It is well known that $|\mu(\boldsymbol{\rho'_{1}},0)|^{2}$ is is often used to characterize the second order statistical properties of the thermal light fields. Their 3D plots are listed in Fig. \ref{fig:2} (\emph{e}) - (\emph{h}). Among them, $\rho'_{1x}$ and $\rho'_{1y}$ are rectangular components for $\boldsymbol{\rho'_1}$ with unit of pixel. In computer experiment, the distance for each pixel is $1.557$ $\mu$m. From Fig. \ref{fig:2} (\emph{e}) - (\emph{h}) one can see the shapes of them are the Fourier transforms of the intensity distribution at the source plane of $\boldsymbol{\rho_{0}}$, and their widths of half maxims are with order of Eq. \ref{eq:16}. They are consistent with predictions by VCZ theorem.

\section{\label{sec:4}Computer experiments and results}
In the computer experiments, we use a double slit as the object, its transmission function is modulated as:
\begin{align}
{\boldsymbol{t}}\left( {\boldsymbol{\rho'_1}} \right) = \operatorname{rect} \frac{{\boldsymbol{\rho'_1} + {\raise0.7ex\hbox{$b$} \!\mathord{\left/
 {\vphantom {b 2}}\right.\kern-\nulldelimiterspace}
\!\lower0.7ex\hbox{$2$}}}}{a} + \operatorname{rect} \frac{{\boldsymbol{\rho'_1} - {\raise0.7ex\hbox{$b$} \!\mathord{\left/
 {\vphantom {b 2}}\right.\kern-\nulldelimiterspace}
\!\lower0.7ex\hbox{$2$}}}}{a}.
\label{eq:18}
\end{align}
For simplicity, only one transverse dimension is considered, although the generalization for two transverse is straightforward. In Eq. \ref{eq:18}, $a = 105$ $\mu$m, is the width for each slit, and $b= 303$ $\mu$m, is the separation distance for two slits. In this work, we regard $a$ as the object's feature size because it describes the smallest feature of the object. The distance parameters $d$, $d_{1}$, and $d_{2}$ in the setup (Fig. \ref{fig:1}) are arranged to be $135$ mm, $60$ mm, and $75$ mm to fulfill Eq. \ref{eq:2}. The parameter $\sigma^{2}$ in Eq. \ref{eq:7} is chosen to be unit. In the following subsections, we applied the convergence criterion (Eq. \ref{eq:1}) to report four computer experiments to investigate a factor that impacts on the iteration numbers $N$. First, let us report:

\subsection{\label{sec:leve4.1}The general convergence trend}
The aperture's diameter of the thermal source is set as $\phi = 50$ mm to make the spatial coherence $l_{c}$ of the light illuminating on the object is $0.532 \times\frac{60}{5} = 6.384$ $ \mu$m (Eq. \ref{eq:16}), clearly it is shorter than the feature size of the object, so that the light fields illuminated on the object was chaotic.

\begin{figure*}\includegraphics
[bb=4 0 246 163,scale=1.6]
{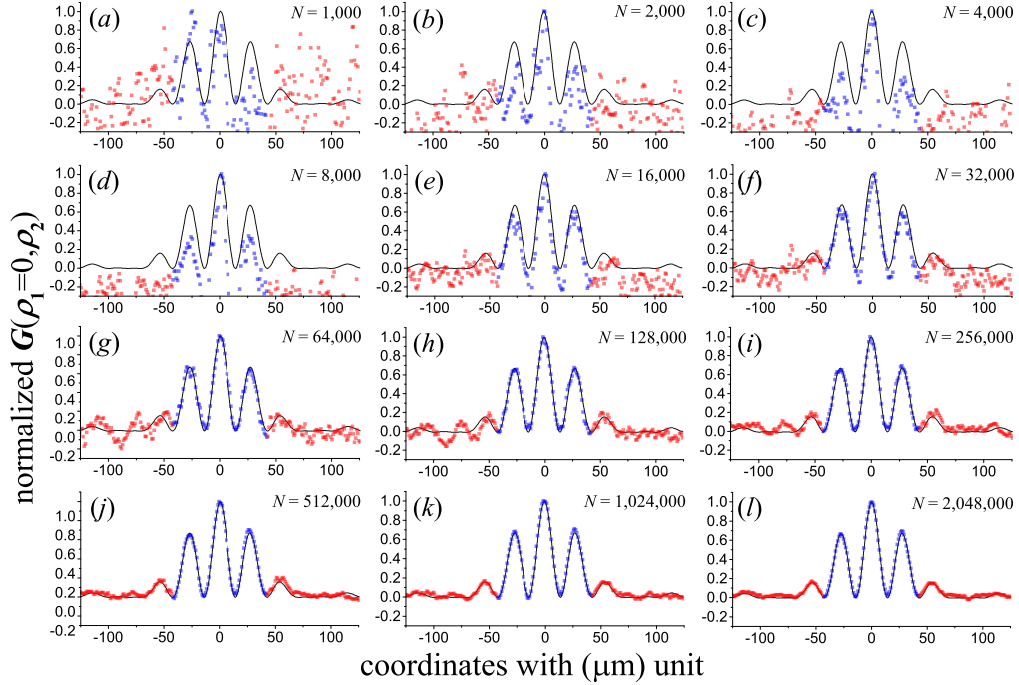}\caption{\label{fig:3} (Color online)  The computer experimental results when $d$, $d_{1}$, and $d_{2}$, are arranged to be $135$ mm, $60$ mm, and $75$ mm. In the experiment, the aperture's diameter $\phi$ is set to $50$ mm. (\emph{a})-(\emph{l}) show the experimental results of $\hat{y}_{i}$ by square dots with $N = 1000\times2^{n}$, $(n = 0, 1,... ,11)$. Among them the continuous lines are the fitted curves for theoretical prediction of $y_{i}$. The experimental data were divided into three parts of equal width which corresponding to, in Fourier space, the low and high frequency components. The low frequency component occupies $\frac{1}{3}$ of the global pattern and the high frequency component occupies $\frac{2}{3}$ of it.}
\end{figure*}

The ghost interference patterns are reconstructed when different numbers $N$ of uncorrelated offline data were used. Fig. \ref{fig:3} (\emph{a})-(\emph{l}) show the experimental results of $\hat{y}_{i}$ by square dots with $N = 1000\times2^{n}$, $(n = 0, 1,\ldots,11)$, respectively. Among them the continuous lines are the fitted curves for theoretical prediction of $y_{i}$. They are the squared modules of the Fourier transform of the object (Eq.\ref{eq:18}) in the form of
\begin{align}
y\left( {\boldsymbol{\rho_2}} \right) = \frac{1}{2}{\operatorname{sinc} ^2}\left( {\frac{a}{{\lambda {d_2}}}\boldsymbol{\rho_2}} \right)\left( {1 + \cos \frac{{2\pi b}}{{\lambda {d_2}}}\boldsymbol{\rho_2}} \right).
\label{eq:19}
\end{align}
By observing through  (\emph{a}) to (\emph{l}) of Fig. \ref{fig:3}, one may appreciate the procedure in which the experimental data was converging.

\begin{figure*}[htb]\includegraphics
[bb=5 85 471 242,scale=0.8]
{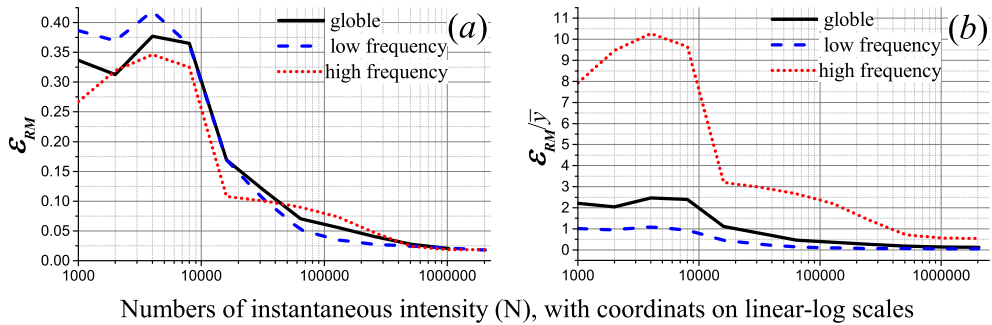}\caption{\label{fig:4} (Color online) (\emph{a}) The convergence trend shown by $\varepsilon_{RM} - N$ for the situations in Fig.\ref{fig:3}(\emph{a})-(\emph{l}).The curve of continuous line illustrates the global convergence trend. The curve of dash and curve of dot illustrate the convergence trends for the low and high frequency components respectively. (\emph{b}) The convergence trend shown by $\frac{\varepsilon_{RM}}{\overline{y}}-N$. The curve of continuous line illustrates the global convergence trend. The curve of dash and curve of dot illustrate the convergence trends for the low and high frequency components respectively. Clearly, in (\emph{b}) their differences are shown more marked than they are in (\emph{a}). }
\end{figure*}

For the convenient of the following investigation, the experimental data for the global pattern were divided into three parts of equal width which corresponding to, in Fourier space, the low and high frequency components. As one can see from Fig. \ref{fig:3}, the low frequency component occupies $\frac{1}{3}$ of the global pattern and the high frequency component occupies $\frac{2}{3}$ of it. By calculating root-mean standard error $\varepsilon_{RM}$ defined by Eq. \ref{eq:1}, their convergence trends was displayed in Fig. \ref{fig:4} (a). In it, the curve of continuous line describes convergence trend for the global patterns; the curve of dash describes the convergence trend for low frequency components; and the curve of dot describes the convergence trend for high frequency components. One can visually see that the global patterns, and the low and high frequency components of them share similar convergent trends in general but with minor differences in detail. If they are compared with relations of $\frac{\varepsilon_{RM}}{\overline{y}}-N$ [Fig. \ref{fig:4} (b)], $\overline{y}$ is the average value for the global patterns and their low and high frequencies components, their differences are more marked.

In the next step, the experiments were performed with changes of illuminating light's spatial coherent length $l_{c}$ to introduce:

\subsection{\label{sec:leve4.2}A linear factor impacts on $N$}
In order to make the results more general, we investigate the factor of the ratio of the object's feature size $a$ to spacial coherent length $l_{c}$,
\begin{align}
\kappa\equiv\frac{a}{l_{c}},
\label{eq:20}
\end{align}
namely feature size relative to the spacial coherent length, that impacts on the convergent speed, with criterion of $\varepsilon_{RM}  = 0.07$ was chosen. In the experiment, $l_{c}$ are modulated according to Eq. \ref{eq:16} by let the value of the $\phi = 1720+20m$ $\mu$m, ($m = 0,1,2,3,4.$), and $2000+200n$ $\mu$m, ($n = 0,1,\ldots,15.$), with the other parameters  unchanged as they were in Sec. \ref{sec:leve4.1}. Under such arrangement, the dependence of $N$
\begin{figure}[htb]\includegraphics
[bb=29 20 468 321,scale=0.50]
{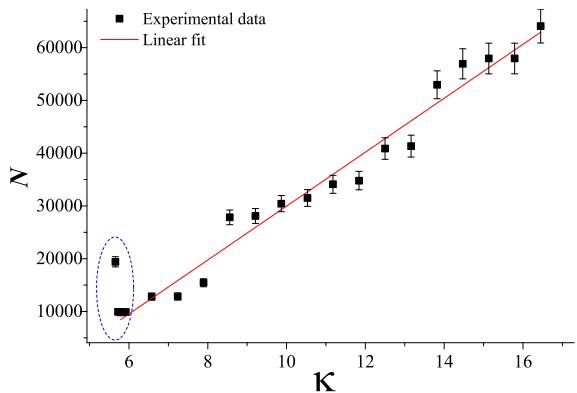}\caption{\label{fig:5} (Color online) The dependence of $N$ on $\kappa$ when $\varepsilon_{RM}  = 0.07$. $\kappa\equiv\frac{a}{l_{c}}$, is the feature size relative to the spacial coherent length of the object. The squared dots are experimental data and the continuous line is their linear fit. It shows when $\kappa>6$, the dependence of $N$ on $\kappa$ fits linear relation well; but when $\kappa<6$, the relation of $N$ and $\kappa$ appeares the none-linear dependence.}
\end{figure}
on $\kappa$ are brought out in Fig. \ref{fig:5}. In it, the squared dots are experimental data and the continuous line is their linear fit. One can see from the main trend that the dependence of $N$ on $\kappa$ fits linear relation well when $\kappa>6$.

Also in Fig. \ref{fig:5}, one may note that the relation of $N$ and $\kappa$  appeared a manner of none-linear dependence (see the oval dotted circle in Fig. \ref{fig:5}) when $\kappa < 6$. Then with all the other parameters used before, we exam this none-linear part by let the value of the $\phi = 1720+4k$ $\mu$m, ($k = 0,1,\ldots,30$), thus lead to $5.66 <\kappa<6.05$. So that we can present:

\subsection{\label{sec:leve4.3}A non-linear factor impacts on $N$}

\begin{figure}\includegraphics
[bb=52 18 438 315,scale=0.65]
{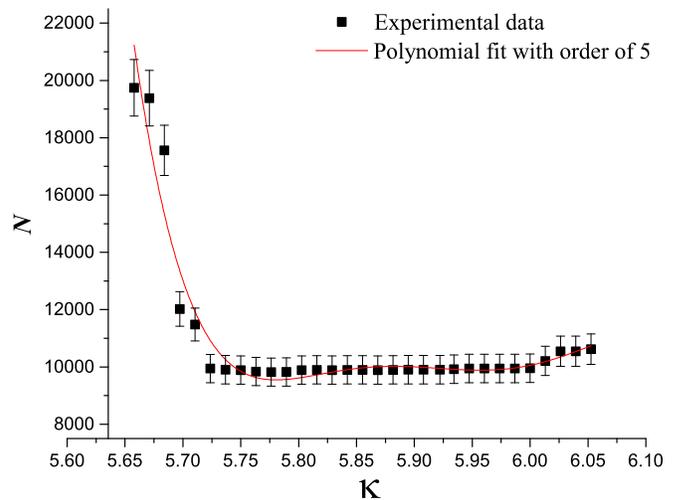}\caption{\label{fig:6} (Color online) The dependence of $N$ on $\kappa$ for the global pattern on the interval of $5.65 <\kappa<6$, when $\varepsilon_{RM} = 0.07$. $\kappa\equiv\frac{a}{l_{c}}$, is the feature size relative to the spacial coherent length of the object.}
\end{figure}

More in detail, the dependence of $N$ on $\kappa$ for the global pattern on the interval of $5.66 <\kappa<6.05$ is shown in Fig. \ref{fig:6}. In the figure, the squared dots are data from the computer experiment, and the continuous line is a polynomial fit with order of $5$. One can see when $5.66 <\kappa< 5.75$, the dependence of $N$ on $\kappa$ is characterized by an opposite trend as it was when
$\kappa>6$ (refer to Fig. \ref{fig:5}). But on an interval of $5.75 <\kappa< 6.05$, the dependence of $N $ on $\kappa$ appears less varied, and thus extended to the linear intervals of $\kappa>6$ (refer to Fig. \ref{fig:5}).

It is worth to note, when perfuming the computer experiment under the condition of $\kappa<5.65$, the criterion of $\varepsilon_{RM }= 0.07$ can not be obtained so far, even with $N = 2,200,000$, the twice order lager than it is for $\kappa = 5.66$. It may be explained if we look a little deeper inside:

\subsection{\label{sec:leve4.4}Opposite convergence tendencies for low and high frequency components}
As we've already seen in Fig. \ref{fig:4}, as $N$ increases, the $\varepsilon_{RM}$'s for the obtained global pattern and for  its low and high frequency components are deferent. Based on this phenomenon, one may
\begin{figure}\includegraphics
[bb=65 18 438 315,scale=0.6]
{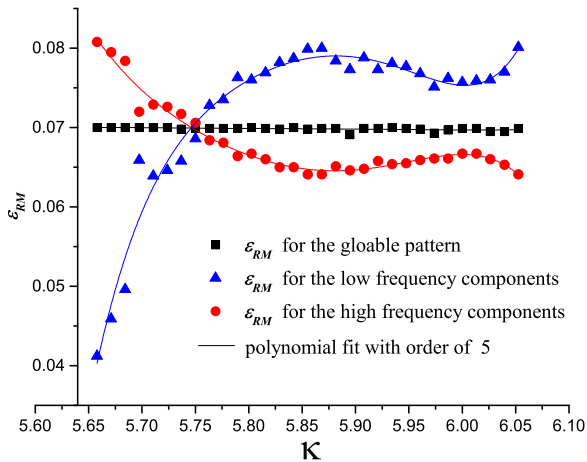}\caption{\label{fig:7} (Color online) The dependence of $\varepsilon_{RM}$'s on $\kappa$ for the low and high components of the obtained patterns when the $\varepsilon_{RM}$ for the global pattern is set to $0.07$. $\kappa\equiv\frac{a}{l_{c}}$, is the feature size relative to the spacial coherent length of the object.}
\end{figure}
naturally assume that if an global pattern reached an appreciable $\varepsilon_{RM}$, say $0.07$, then the $\varepsilon_{RM}$'s for its low and high frequency components are not necessarily the same. By a computer experiment, we'll see this is indeed the case. In Fig. \ref{fig:7} we present the changes of $\varepsilon_{RM}$'s with respect to $\kappa$, for both low and high frequency components while let $\varepsilon_{RM} = 0.07$ for the global pattern. In the figure, the squared dots are the $\varepsilon_{RM}$ for the global pattern, the triangular dots are the $\varepsilon_{RM}$ for the experimental data of low frequency component, and the round dots are the $\varepsilon_{RM}$ for the experimental data of high frequency component. One can see in Fig. \ref{fig:7}, they are different as expected and the convergence trends for low and high frequency components are with opposite tendencies on the interval of 5.65 $<\kappa< 6.05$.

On the interval of $5.65< \kappa < 5.75$, the $\varepsilon_{RM}$ for low frequency component is smaller than it is for the global pattern, while the $\varepsilon_{RM}$ for high frequency component is larger than it is for the global pattern. This implies the low frequency component is more likely to convergent than the high frequency component is. Although so, it is possible for their weighted average to keep $\varepsilon_{RM}$ unchanged at $0.07$ for the global pattern. On the interval of $5.75 < \kappa< 6.05$, the situation is in opposite. The $\varepsilon_{RM}$ for low frequency component is larger than it is for the global pattern, while the $\varepsilon_{RM}$ for high frequency components is smaller than it is for the global pattern. This implies the high frequency components is more likely to convergent than the low frequency components is. Although so, it is also possible for their weighted average to keep $\varepsilon_{RM}$  unchanged at $0.07$ for the global pattern.

As we can see, when $\kappa < 5.65$, while it decreases, although the $\varepsilon_{RM}$ for low frequency components is going to below $0.04$ with a limit no less than zero, the $\varepsilon_{RM}$ for high frequency component is gonging to exceed $0.08$ without a limit. Thus their weighted root-mean standard error, $\varepsilon_{RM}$, for the global pattern is impossible to converge at the criterion of $\varepsilon_{RM} = 0.07$. This feature explains the last situation described in Sec. \ref{sec:leve4.3}.

\section{\label{sec:5}Conclusions}
We introduce a criterion based on the concept known as ``root-mean standard error" $\varepsilon_{RM}$, to valuate the convergent quality of GI.

Based on this criterion, we performed four computer experiments. The first one reproduced a common sense for the procedure that the experimental data converges to its theoretical prediction as $N$ increases. If we divided the global pattern into low and high frequency components, we find they have different convergent speeds (Sec. \ref{sec:leve4.1}).

The second experiment investigated a factor of $\kappa$, (Eq. \ref{eq:20}),  namely feature size relative to the spacial coherent length that impacts on the convergent speed. It is found, when $\kappa>6$, the dependence of the necessary number of uncorrelated data acquired for the correlation computation, $N$, on $\kappa$ fits linear relation well (Sec. \ref{sec:leve4.2}). On this interval, the larger $\kappa$ is, the larger number of $N$ is required for the experimental data to converge.

In the third experiment, We observed that $\kappa$ acted like a non-linear factor that impacts on $N$. We found, on the interval of  $[5.66,5.75]$, the dependence of $N$ on $\kappa$ is characterized by an opposite trend as it was when $\kappa > 6$, i.e. in this interval, the larger $\kappa$ is, the smaller number of $N$ is required for the experimental data to converge. We also found, when $5.75<\kappa<6.0$, the dependence of $N$ on $\kappa$ appears less varied (Sec. \ref{sec:leve4.3}). These facts suggest, when performing the experiment, one needs to control the illuminations' spatial coherent length $l_{c}$ to make $\kappa$ to fall into an interval of $(5.75, 6.0)$, thus to yield a satisfied precisely convergent GI with a minimal $N$.

We observed, in the last experiment, there exists opposite convergence tendencies for low and high frequency components when $5.65< \kappa< 6.05$ (Sec. \ref{sec:leve4.4}). This observation suggests that one needs to select a preferred $\kappa$ based on the spatial frequency range he was interested in, in order to accelerate the imaging speed. It is worth to note, when $\kappa < 5.65$, the experimental data for global pattern is not able to obtain the criterion of $\varepsilon_{RM} = 0.07$.

In addition, the method to simulate the uncorrelated sequence of a complex ensemble for thermal light (which was introduced in Sec. \ref{sec:3}) is valuable to actively control the light source for the real experiment, e.g. for the computational imaging with single-pixel \cite{17,20,21,23}.

\begin{acknowledgments}
The author would like to thank Professor Ming Yang for the computational facilities supported by NSFC under Grant No.11274010.
\end{acknowledgments}

\end{document}